\documentclass[conference,10pt,epsfig]{IEEEtran}

\usepackage[colorlinks=true]{hyperref}
\usepackage{cite}
\usepackage{amsmath,amssymb,amsfonts}
\newtheorem{proposition}{Proposition}

\usepackage{algorithmic}
\usepackage{graphicx}
\usepackage{textcomp}
\usepackage{xcolor}
\usepackage{cleveref}
\usepackage{geometry}
\usepackage{subcaption}
\usepackage{algorithm}
\usepackage{algorithmic}
\usepackage{siunitx}
\usepackage{booktabs}
\usepackage{multirow}
\usepackage{array}
\usepackage[font=small]{caption}
\def\BibTeX{{\rm B\kern-.05em{\sc i\kern-.025em b}\kern-.08em
    T\kern-.1667em\lower.7ex\hbox{E}\kern-.125emX}}

\begin{document}
\title{Rotatable Antenna Array-Enhanced Null Steering: Performance Analysis and Optimization}
\IEEEoverridecommandlockouts
\author{
\IEEEauthorblockN{
    Yingqi Wen\IEEEauthorrefmark{1}, 
    Weidong Mei\IEEEauthorrefmark{1}, 
    Yike Xie\IEEEauthorrefmark{1}, 
    Beixiong Zheng\IEEEauthorrefmark{2},
    Zhi Chen\IEEEauthorrefmark{1}, 
    and Boyu Ning\IEEEauthorrefmark{1} 
}
\IEEEauthorblockA{
    \IEEEauthorrefmark{1}National Key Laboratory of Wireless Communications,\\
    University of Electronic Science and Technology of China, Chengdu 611731, China\\
}
\IEEEauthorblockA{
    \IEEEauthorrefmark{2}School of Microelectronics, South China University of Technology, Guangzhou 510641, China\\
    Emails: 2023190504033@std.uestc.edu.cn; wmei@uestc.edu.cn; ykxie@std.uestc.edu.cn;\\ bxzheng@scut.edu.cn; chenzhi@uestc.edu.cn; boydning@outlook.com;}
}
\maketitle

\begin{abstract}
Conventional fixed-orientation antenna (FOA) arrays offer limited degrees of freedom (DoF) for flexible beamforming such as null steering. To address this limitation, we propose a new rotatable antenna array (RAA) architecture in this paper, which enables three-dimensional (3D) rotational control of an antenna array to provide enhanced spatial flexibility for null steering. To characterize its performance, we aim to jointly optimize the 3D rotational angles of the RAA, to maximize the beam gain over a given desired direction, while nulling those over multiple interference directions under zero-forcing (ZF) beamforming. However, this problem is non-convex and challenging to tackle due to the highly nonlinear expression of the beam gain in terms of the rotational angles. To gain insights, we first examine several special cases including both isotropic and directional antenna radiation patterns, deriving the conditions under which full beam gain can be achieved over the desired direction while meeting the nulling constraints for interference directions. These conditions clearly indicate that compared with FOA arrays, RAAs can significantly relax the angular separation requirement for achieving effective null steering. For other general cases, we propose a sequential update algorithm, that iteratively refines the 3D rotational angles by discretizing the 3D angular search space. To avoid undesired local optimum, a Gibbs sampling (GS) procedure is also employed between two consecutive rounds of sequential update for solution exploration. Simulation results verify our analytical results and show superior null-steering performance of RAAs to FOA arrays.
\end{abstract}

\section{Introduction}
Beamforming is a fundamental technique in multi-antenna systems, enabling directional transmission and interference suppression by coherently adjusting the phase and/or amplitude of antenna elements. Over the past decades, it has become a key technology in wireless communications, radar, and sensing applications\cite{b1,b2}. However, conventional fixed-position and orientation arrays inherently suffer from fixed spatial correlation among steering vectors, which imposes a fundamental trade-off between maximizing beamforming gain in a desired direction and minimizing the beamforming gain over interference directions \cite{b1,b2}. Classical beamforming methods, such as zero-forcing (ZF) beamforming, often result in beam-gain loss at the desired direction, especially if it is close to one or multiple interference directions.

To mitigate this issue, movable antenna (MA) technology has been proposed to reconcile this trade-off by adjusting antenna positions and geometry in a confined region, thereby reconfiguring the spatial correlation among the steering vectors corresponding to different angles\cite{b5, b6, b6a}. It has been shown in \cite{b7} that by properly optimizing the MA positions, an MA array can maximize the beam gain over the desired direction and null those over multiple interference directions at the same time, referred to as {\it steering vector orthogonality} (SVO) in this paper. However, an MA array generally requires increasing the antenna aperture to achieve the SVO, which may face difficulty in practice \cite{b7}. More recently, rotatable antenna technology has been proposed as a more efficient alternative to MAs in the literature \cite{b8, b9, b10, b11}. Compared with MA arrays, rotatable antenna arrays (RAAs) control only the orientation of each antenna element or the entire array to reshape the antenna radiation pattern, thereby reducing implementation complexity. Although RAAs have been previously studied in various system setups \cite{b8, b9, b10, b11}, most of them focus on multi-path channel setups. Hence, there is still a lack of in-depth performance analysis and optimization of RAAs in the beam domain from an array signal processing perspective.
\begin{figure}
    \centering
    \includegraphics[width=1.0\linewidth]{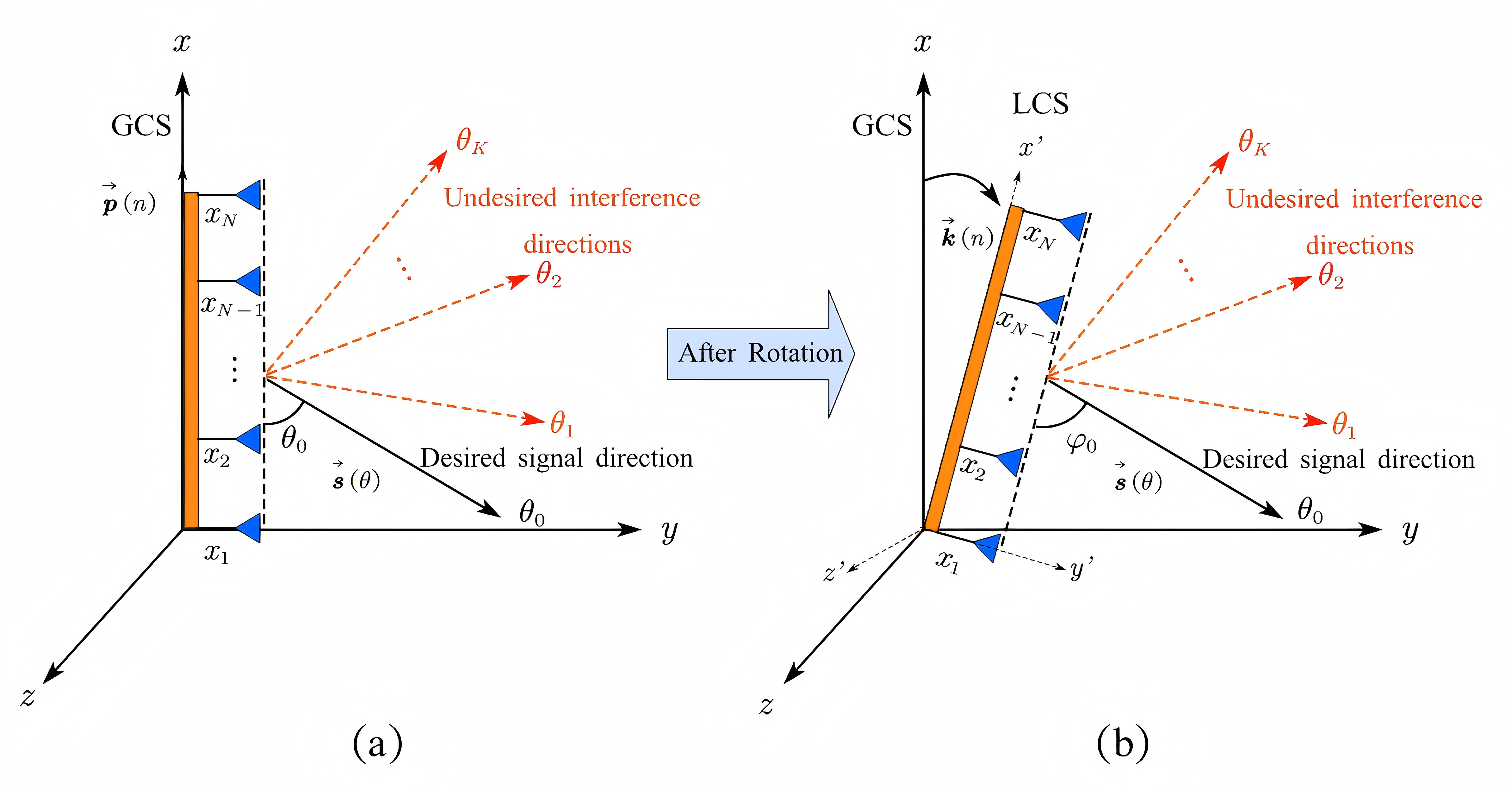}
    \caption{Illustration of an RAA in GCS and LCS.}
    \vspace{-12pt}
    \label{fig:array_rotation}
\end{figure}

Motivated by the above, we investigate the use of RAAs for array signal processing and focus on null steering in this paper, as shown in Fig.~\ref{fig:array_rotation}. Particularly, we aim to jointly optimize the three-dimensional (3D) rotational angles of the RAA to maximize the beam gain over a desired direction while strictly nulling multiple interference directions under zero-forcing (ZF) beamforming. However, due to the highly nonlinear structure of the beam gain with respect to the rotational angles, this problem is inherently non-convex and challenging to be optimally solved. To gain insights, we first analyze several special cases including both isotropic and directional antenna radiation patterns, and derive their corresponding conditions for achieving SVO. These results reveal that RAA can effectively relax the angular separation requirements compared with conventional fixed-orientation antenna (FOA) array for null steering. For general scenarios, we develop a sequential update algorithm to obtain suboptimal 3D rotational angles iteratively, jointly with Gibbs sampling to avoid local optimum via solution exploration. Simulation results verify our analytical results and demonstrate the superior performance of RAAs to FOA arrays under various setups.

\begingroup
\allowdisplaybreaks
\section{System Model}
As shown in Fig.~\ref{fig:array_rotation}, we consider a rotatable uniform linear antenna array (ULA) with $N$ elements, which have an equal spacing denoted as $d$. For convenience, we establish a global coordinate system (GCS) and assume that the initial direction of the rotatable ULA is parallel to the $x$-axis, or equivalently, its initial boresight direction is parallel to the $y$-axis, as shown in Fig.~\ref{fig:array_rotation}(a). Hence, the initial positions of the $N$ antennas are given by $\mathbf{p}(n) = [nd, 0, 0]^T, \; n=0,1,\dots,N-1$, and the initial boresight direction of the ULA is given by $\mathbf{f} = [0,1,0]^T$.

We consider that the rotatable ULA can be flexibly rotated along the $x$-, $y$-, and $z$-axes in the 3D space by employing a gimbal. To describe the array rotation, we also define a local coordinate system (LCS) for the ULA after its rotation, which is assumed to be parallel to $x'$-axis, as shown in Fig.~\ref{fig:array_rotation}(b). The array rotation is represented by an array rotation vector (ARV) $\mathbf{r} = [\alpha, \beta, \gamma]^T$, where $\alpha$, $\beta$, and $\gamma \in [0, 2\pi]$ denote the rotational angles around the $x$-, $y$-, and $z$-axes, respectively. The transformation from LCS to GCS can be characterized by a rotation matrix
\begin{equation}
\mathbf{R} = \begin{bmatrix}
c_\beta c_\gamma & s_\alpha s_\beta c_\gamma - c_\alpha s_\gamma & c_\alpha s_\beta c_\gamma + s_\alpha s_\gamma \\
c_\beta s_\gamma & s_\alpha s_\beta s_\gamma + c_\alpha c_\gamma & c_\alpha s_\beta s_\gamma - s_\alpha c_\gamma \\
-s_\beta & s_\alpha c_\beta & c_\alpha c_\beta
\end{bmatrix}.
\label{eq:rotation_matirx}
\end{equation}
where $c_\psi = \cos(\psi)$ and $s_\psi = \sin(\psi)$, $\psi \in \{\alpha, \beta, \gamma\}$. Based on (\ref{eq:rotation_matirx}), the coordinate of the $n$-th antenna in the LCS is
\begin{equation}
    \mathbf{k}(n) = \mathbf{R}\mathbf{p}(n) = [c_\beta c_\gamma, \; c_\beta s_\gamma, \; -s_\beta]^T nd,
    \label{rotated_array}
\end{equation}
and the boresight direction after the rotation is given by
\begin{equation}
    \mathbf{f}'(n) = \mathbf{R}\mathbf{f}(n) 
    = \big[ -c_\alpha s_\gamma + s_\alpha s_\beta c_\gamma, \; 
            c_\alpha c_\gamma + s_\alpha s_\beta s_\gamma, \;
            s_\alpha c_\beta \big]^T.
\end{equation}

\begin{figure}
    \centering
    \includegraphics[width=0.6\linewidth]{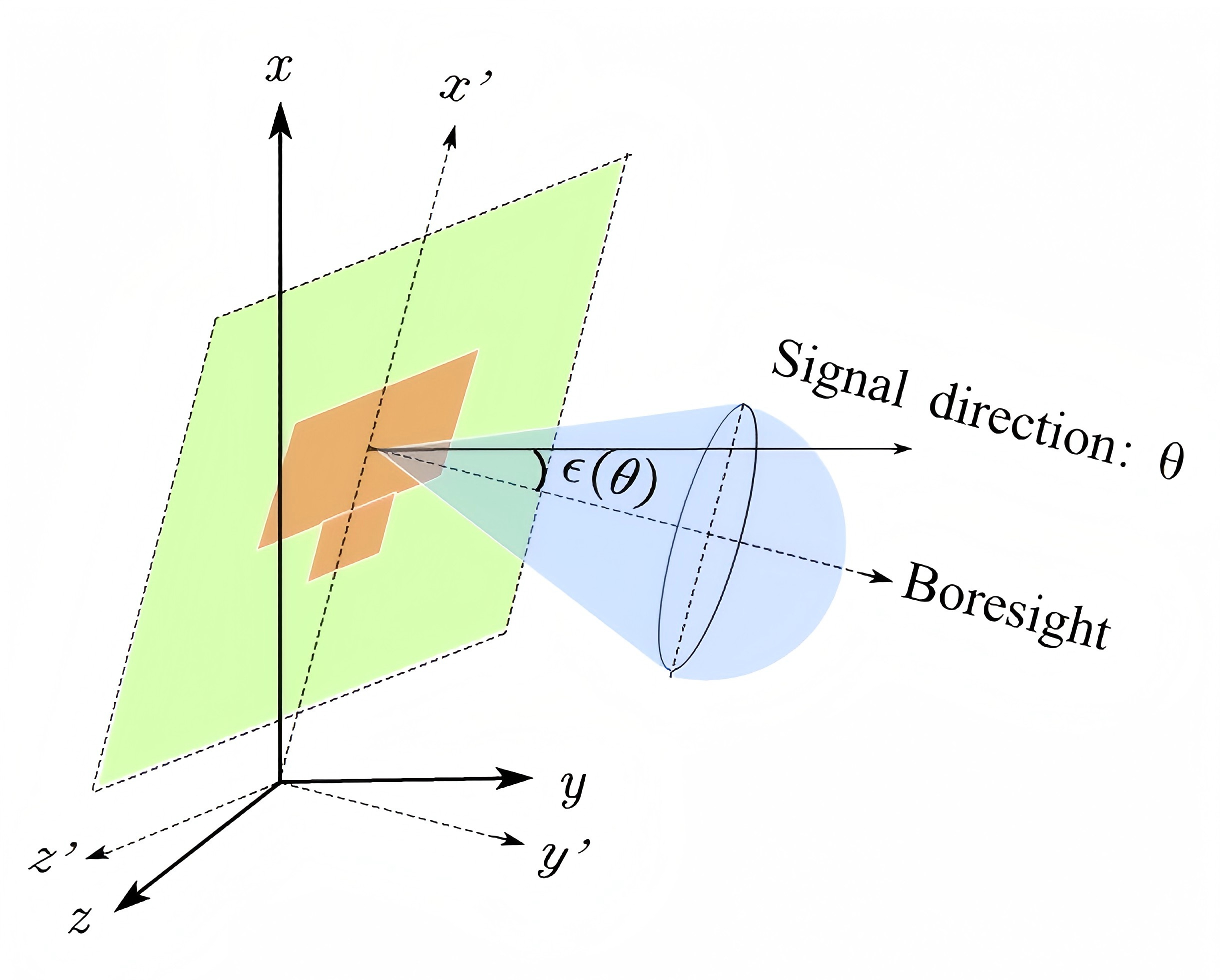}
    \caption{Illustration of array rotation and boresight direction.}
    \vspace{-12pt}
    \label{Angular_deviation}
\end{figure}
Let $\theta$ and $\varphi$ denote the angles of departure (AoDs) before and after the array rotation, respectively, as shown in Fig.  \ref{fig:array_rotation}. Then, the signal direction vector can be expressed as $\mathbf{s}(\theta)=[-\cos\theta, \sin\theta, 0]^T$. Based on geometry, the AoD after the rotation can be obtained based on the angle between the rotated array and the signal direction, i.e.,
\begin{equation}
    \cos \varphi = \frac{\mathbf{k}(n)}{|\mathbf{k}(n)|} \cdot \mathbf{s}(\theta) 
    = -\cos\beta \cos(\gamma + \theta).
    \label{eq:sol_fai}
\end{equation}
It is noted from (\ref{eq:sol_fai}) that the AoD after the rotation depends on both $\beta$ and $\gamma$, but is independent of $\alpha$. Based on (\ref{eq:sol_fai}), the array response of the $n$-th antenna after the rotation is expressed as
\begin{equation}
    a_n(\mathbf{r}, \theta) = e^{j \frac{2\pi}{\lambda} n d \cos \varphi} 
    = e^{-j \frac{2\pi}{\lambda} nd \cos \beta \cos(\gamma+\theta)}.
    \label{no_gain_response}
\end{equation}
The corresponding geometric steering vector is written as
\begin{equation}
    {\mathbf{a}}(\mathbf{r}, \theta) = [{a}_0(\mathbf{r}, \theta), \; {a}_1(\mathbf{r}, \theta), \; \dots, \; {a}_{N-1}(\mathbf{r}, \theta)]^T.
    \label{eq:ini_Sv}
\end{equation}

For each antenna element, it may have a non-isotropic radiation pattern. Assume that all antennas share the same radiation pattern denoted as $g(\mathbf{r}, \epsilon(\theta))\geq 0$, where $\epsilon(\theta)$ denotes the angle between the boresight of each antenna element after rotation and the signal direction vector (i.e., $\mathbf{s}(\theta)$) and satisfies
\begin{equation}
    \cos (\epsilon(\theta))= \mathbf{f}'(n) \cdot \mathbf{s}(\theta) 
    = -s_\alpha s_\beta \cos(\gamma+\theta) + c_\alpha s_\gamma(\gamma+\theta).
    \label{eq:sol_ep}
\end{equation}
Notably, all elements should share the same boresight direction and angle $\epsilon(\theta)$. By incorporating the antenna radiation pattern, the steering vector in (\ref{eq:ini_Sv}) becomes
\begin{equation}
    \tilde{\mathbf{a}}(\mathbf{r}, \theta) = \sqrt{g(\mathbf{r}, \epsilon(\theta))} \mathbf{a}(\mathbf{r}, \theta),
    \label{effective_SV}
\end{equation}
which is referred to as effective steering vector in this paper. Note that unlike the geometric steering vector in (\ref{eq:ini_Sv}), the effective steering vector in (\ref{effective_SV}) also depends on $\alpha$.

Finally, let $\mathbf{w} \in \mathbb{C}^N$ denote the beamforming vector. For any given rotation angles $(\alpha, \beta,\gamma)$, the beam gain at $\theta$ is given by
\begin{equation}
    G(\mathbf{r},\mathbf{w},\theta) = \left| \tilde{\mathbf{a}}(\mathbf{r}, \theta)^H \mathbf{w}\right|^2.
    \label{eq:beam_gain}
\end{equation}
It is noted from (\ref{eq:beam_gain}) that by optimizing $\mathbf{r}$, the beam gains over different directions can be flexibly adjusted, in addition to conventional beamforming designs.

\section{Problem Formulation}
This paper focuses on an interference-dominant scenario, with a goal to maximize the beam gain over a desired angle (denoted as $\theta_0$) while nulling $K$ interference directions (denoted as $\theta_k,k\in {\cal K} \triangleq \{1, 2,\cdots, K\}$) by jointly optimizing the ARV $\mathbf{r}$ and the beamforming vector $\mathbf{w}$. The associated optimization problem can be formulated as
\begin{align}
\textrm{(P1)}\;\max_{\mathbf{r},\mathbf{w}} \quad & G(\mathbf{r},\mathbf{w}, \theta_0) \label{12a}\\
\text{s.t.} \quad & G(\mathbf{r},\mathbf{w},\theta_k)=0,\quad k \in \cal K, \label{12b}\\
& \alpha, \beta, \gamma \in [0,2\pi), \label{12c}\\
& \|\mathbf{w}\|_2^2 = 1, \label{12d}
\end{align}
where \eqref{12b} is the null-steering condition, while \eqref{12d} normalizes the transmit beamforming.

It is known that for any given ARV $\mathbf{r}$, the optimal beamforming vector to maximize $G(\mathbf{r},\mathbf{w}, \theta_0)$ under the null-steering constraints \eqref{12b} is given by ZF beamforming, i.e.,
\begin{align}
\mathbf{w}^{\mathrm{ZF}}({\bf{r}}) &= 
{\mathbf{w}({\bf{r}})}/{\|\mathbf{w}({\bf{r}})\|_2}, \nonumber\\
\mathbf{w}({\bf{r}}) &= 
\left[\mathbf{I} 
- \tilde{\mathbf{A}}(\mathbf{r})
\big(\tilde{\mathbf{A}}(\mathbf{r})^{H}
\tilde{\mathbf{A}}(\mathbf{r})\big)^{-1}
\tilde{\mathbf{A}}^{H}(\mathbf{r}) \right]
\tilde{\mathbf{a}}(\mathbf{r}, \theta_0),
\label{eq:ZF_AWV}
\end{align}
where $\tilde{\mathbf{A}}(\mathbf{r}) = [\tilde{\mathbf{a}}(\mathbf{r},\theta_1), \tilde{\mathbf{a}}(\mathbf{r},\theta_2),\cdots,  \tilde{\mathbf{a}}(\mathbf{r},\theta_K)]$ is the interference-steering matrix. By separating the steering vector and antenna radiation pattern, this matrix can be rewritten as
\begin{equation}
    \tilde{\mathbf{A}}(\mathbf{r}) = [\mathbf{a}(\mathbf{r}, \theta_1), \mathbf{a}(\mathbf{r}, \theta_2), \dots, \mathbf{a}(\mathbf{r}, \theta_K)] \cdot \mathbf{D}
    = \mathbf{A}(\mathbf{r}) \cdot \mathbf{D},
    \label{eq:SV_interf}
\end{equation}
where $\mathbf{A}(\mathbf{r}) = [ \mathbf{a}(\mathbf{r},\theta_1),  \mathbf{a}(\mathbf{r},\theta_2), \dots,  \mathbf{a}(\mathbf{r},\theta_K)]$ collects geometric steering vectors, and the diagonal matrix $\mathbf{D} = \mathrm{diag}(\sqrt{g(\mathbf{r}, \epsilon(\theta_1))}, \sqrt{g(\mathbf{r}, \epsilon(\theta_2))}, \dots, \sqrt{g(\mathbf{r}, \epsilon(\theta_K))})$ accounts for the antenna radiation patterns in all interference directions. It is interesting to note that if $g(\mathbf{r}, \epsilon(\theta))=0$, then the beam gain at $\theta$ can be naturally nulled without the need for ZF beamforming. By substituting (\ref{eq:SV_interf}) into (\ref{eq:ZF_AWV}), we can obtain
\begin{align}
\mathbf{w}({\bf{r}})=&\nonumber \\
\big[\mathbf{I}& - \mathbf{A}(\mathbf{r})\mathbf{D}\mathbf{D}^{-1}(\mathbf{A}(\mathbf{r})^H \mathbf{A}(\mathbf{r}))^{-1}\mathbf{D}^{-H}\mathbf{D}\mathbf{A}(\mathbf{r})^H \big] \tilde{\mathbf{a}}(\mathbf{r}, \theta_0) \nonumber \\
    =& \big[\mathbf{I} - \mathbf{A}(\mathbf{r})(\mathbf{A}(\mathbf{r})^H \mathbf{A}(\mathbf{r}))^{-1}\mathbf{A}(\mathbf{r})^H \big] \tilde{\mathbf{a}}(\mathbf{r}, \theta_0) \nonumber \\
    =&\mathbf{P}_\perp  \tilde{\mathbf{a}}(\mathbf{r}, \theta_0),
\end{align}
where $\mathbf{P}_\perp = \mathbf{I} - \mathbf{A}(\mathbf{r})(\mathbf{A}(\mathbf{r})^H \mathbf{A}(\mathbf{r}))^{-1} \mathbf{A}(\mathbf{r})^H$ is termed as the projection operator. It is noted that $\mathbf{D}$ cancels out in $\mathbf{P}_\perp$, implying that the nulling operation depends only on the geometric steering vectors. As a result, the beam gain over the desired direction is given by
\begin{align}
G&(\mathbf{r},\mathbf{w}^{\mathrm{ZF}}({\bf{r}}),\theta_0)\nonumber\\
    &=\|\mathbf{P}_\perp \tilde{\mathbf{a}}(\mathbf{r}, \theta_0)\|_2^2 \\
    &= g(\mathbf{r}, \epsilon(\theta_0)) \|\mathbf{P}_\perp \mathbf{a}(\mathbf{r}, \theta_0)\|_2^2 \\
    &=\! g(\mathbf{r},\!\epsilon(\theta_0))\big(N \!-\! \mathbf{a}(\mathbf{r}, \theta_0)^H \mathbf{A} \big(\mathbf{A}^H \mathbf{A}\big)^{-1} \mathbf{A}^H  \mathbf{a}(\mathbf{r}, \theta_0)\big).\label{eq:ZF_gain}
\end{align}
Evidently, the ZF-induced beam gain in \eqref{eq:ZF_gain} is upper-bounded by the full beam gain $Ng_0$, where $g_0$ denotes the maximum achievable antenna gain for each element. Based on \eqref{eq:ZF_gain}, problem (P1) can be equivalently reformulated as
\begin{equation}
    \textrm{(P2)}\;\max_{\alpha, \beta,\gamma \in [0,2\pi)} \; G(\mathbf{r},\mathbf{w}^{\mathrm{ZF}}({\bf{r}}),\theta_0).
    \label{P2}
\end{equation}

However, (P2) is a non-convex optimization problem that is difficult to be optimally solved, particularly due to the highly nonlinear antenna radiation pattern with respect to the ARV. To gain insights into array rotation for null steering, we consider the following two specific models of antenna radiation patterns and conduct performance analyses accordingly in Section \ref{sp}. Specifically, the first model assumes isotopic antennas, where each antenna element has a uniform beam gain towards all directions, i.e.,
\begin{equation}\label{eq:iso_gain_fac}
    g(\mathbf{r}, \epsilon(\theta))=g_0=1,\quad \forall \theta.
\end{equation}

The second model assumes a directional antenna with its antenna radiation pattern given by \cite{b9}
\begin{equation}
 g(\mathbf{r}, \epsilon(\theta))=
\begin{cases}
    g_0 \cos^{2p} \epsilon(\theta), & \epsilon(\theta) \in \left[0, \tfrac{\pi}{2}\right], \\
    0, & \text{otherwise},
\end{cases}
\label{eq:pattern}
\end{equation}
where $p > 0$ is the directivity factor that determines the beamwidth of the main lobe, and $g_0 = 2(2p+1)$ is the maximum gain in the boresight direction (i.e., $\epsilon = 0$), ensuring power conservation. Note that the radiation model in (\ref{eq:pattern}) can form a single narrow beam within $[0, \pi/2]$ as its mainlobe and negligible sidelobes outside this interval.

\section{Special Case Analysis}\label{sp}
In this section, we investigate several special cases based on the above two radiation pattern models for driving insights. Particularly, to circumvent the difficulty in solving (P2), we focus on deriving the ARV solution that satisfies $\mathbf{A}^H\mathbf{a}(\mathbf{r}, \theta_0)=0$, such that the beam gain in (\ref{eq:ZF_gain}) attains its maximum value of $Ng_0$. This is equivalent to achieving the following effective SVO, i.e.,
\begin{equation}
    {\tilde{\mathbf{a}}}(\mathbf{r}, \theta_k)^H  {\tilde{\mathbf{a}}}( \mathbf{r}, \theta_0) = 0, \quad k \in {\cal K}.
\label{orthog_mei}
\end{equation}
\subsection{Isotropic Antenna}
For isotropic antennas, we can simplify (\ref{orthog_mei}) as 
\begin{equation}
    \mathbf{a}(\mathbf{r},\theta_k)^H \mathbf{a}(\mathbf{r},\theta_0) = 0,
    \quad
    k \in \cal K.
    \label{eq:mul_orthog}
\end{equation}
Based on (\ref{no_gain_response}), (\ref{eq:mul_orthog}) is only related to two-dimensional (2D) rotational angles $\beta$ and $\gamma$. To facilitate our analysis, we next consider the following two cases with $K=1$ and $K\geq 2$, respectively.

\subsubsection{Single Interference Direction ($K=1$)}
In the case of a single interference direction $\theta_1$, the equality in (\ref{eq:mul_orthog}) further reduces to
\begin{equation}
    \mathbf{a}(\mathbf{r}, \theta_1)^H \mathbf{a}(\mathbf{r}, \theta_0) = 0.
\label{eq:orthog}
\end{equation}
The following proposition provides the sufficient and necessary condition under which (\ref{eq:orthog}) holds.
\begin{proposition}\label{prop1}
For any RAA with $N>1$ antennas, there always exist a pair of feasible angles $\beta$ and $ \gamma$ satisfying (\ref{eq:orthog}), if and only if the angular separation between the desired and interference directions satisfies 
\begin{equation}
    |\theta_0 - \theta_1| \in \left[2\arcsin \tfrac{\lambda}{2Nd}, \; 2\pi - 2\arcsin \tfrac{\lambda}{2Nd}\right].
    \label{range_iso_1}
\end{equation}
\end{proposition}
\begin{IEEEproof}
Note that the equality in (\ref{eq:orthog}) can be expanded as
\begin{equation}
    \sum_{n=0}^{N-1} e^{j \frac{2\pi}{\lambda} n d (\cos \varphi_0 - \cos \varphi_1)} = 0.
    \label{eq:geo_series}
\end{equation}
As the left-hand side of (\ref{eq:geo_series}) is a geometric series, it can be calculated via Dirichlet kernel as
\begin{equation}
    e^{j \frac{(N-1)\pi d}{\lambda}\Delta_1} 
    \frac{\sin \big(\tfrac{N\pi d}{\lambda}\Delta_1\big)}{\sin \big(\tfrac{\pi d}{\lambda}\Delta_1\big)} = 0,
    \label{kernel}
\end{equation}
where $\Delta_1 = \cos \varphi_0 - \cos \varphi_1$. The equality in (\ref{kernel}) holds if and only if
\begin{equation}
    \Delta_1 = {m_1\lambda}/{Nd}, 
    \qquad m_1 \in \mathbb{Z}\setminus \tilde{\mathbb{Z}},
    \label{m_condition}
\end{equation}
where $\tilde{\mathbb{Z}} = \{n | n=Ns, s \in {\mathbb{Z}}\}$, i.e., the set of all integers divisible by $N$. By substituting (\ref{eq:sol_fai}) into (\ref{m_condition}), we can obtain the solution for $\beta$ with respect to $\gamma$ as
\begin{equation}
    \cos\beta = \frac{{m_1\lambda}}{2Nd \sin \big({\gamma+(\theta_0+\theta_1)}/{2}\big) \sin \big(({\theta_0 - \theta_1})/{2}\big)} .
    \label{eq:cons_beta}
\end{equation}

However, the equality in (\ref{eq:cons_beta}) is only feasible if its right-hand side lies within $[-1, 1]$, which leads to
\begin{equation}
    \left|\sin(\gamma+\frac{\theta_0+\theta_1}2)\right|\geq \left|\frac{{m_1\lambda}}{2Nd \sin \big(({\theta_0 - \theta_1})/{2}\big)} \right|.
\label{eq:sin_gamma}
\end{equation}

For (\ref{eq:sin_gamma}) to hold, its right-hand side should be no larger than one, leading to
\begin{equation}
    |m_1| \leq ({2Nd}/{\lambda}) \cdot \left|\sin (({\theta_0 - \theta_1})/{2})\right|.
    \label{eq:cons_m}
\end{equation}
Finally, since $|m_1|\geq 1$ for $N>1$, a feasible $m_1$ exists if and only if the right-hand side of (\ref{eq:cons_m}) is no smaller than 1, which leads to (\ref{range_iso_1}). Under this condition, we can retrieve the corresponding $\gamma$ and $\beta$ based on (\ref{eq:sin_gamma}) and (\ref{eq:cons_beta}), respectively.
\end{IEEEproof}

Proposition \ref{prop1} reveals that the feasibility of (\ref{eq:orthog}) depends on the angular separation between the desired and interference directions. Moreover, as $N$ increases, the interval in (\ref{range_iso_1}) becomes enlarged, rendering the SVO condition in (\ref{range_iso_1}) easier to meet. For performance comparison, we further consider the conventional FOA array, for which the orthogonality condition can be obtained as
\begin{equation}
    \cos \theta_0 - \cos \theta_1 = {m_0\lambda}/{Nd}, 
    \qquad m_0 \in \mathbb{Z}\setminus \tilde{\mathbb{Z}}.
    \label{m_cond_dis}
\end{equation}
It follows that (\ref{eq:orthog}) holds for fixed ULA only at a finite set of discrete angles. Compared with the FOA array, the RAA provides additional degree of freedom to align the AoD difference, i.e., ($\cos \varphi_0 - \cos \varphi_1$) with one of these discrete points by optimizing $\beta$ and $\gamma$, thereby extending the feasible region for (\ref{eq:orthog}) from discrete to continuous. This also indicates that array rotation can bring performance gains over FOA arrays even under isotropic antenna radiation patterns.

\subsubsection{Multiple Interference Directions ($K\geq2$)}
Given multiple interference directions $\theta_i, i \in \cal K$, the following proposition shows the conditions for (\ref{eq:mul_orthog}) to hold.
\begin{proposition}\label{prop2}
Let
\begin{equation}
    \begin{split}
        M_{1i} &= m_1 S_i \cos e_i - m_i S_1 \cos e_1, \\
        N_{1i} &= m_1 S_i \sin e_i - m_i S_1 \sin e_1,
    \end{split}
\end{equation}
where $e_i = ({\theta_0 + \theta_i})/{2},$ $S_i = \sin\big(({\theta_0 - \theta_i})/{2}\big)$. Then, for any RAA with $N>1$, if  there exists a constant ratio $\eta$ such that
\begin{equation}
    \eta=\frac{M_{1i}}{N_{1i}}=\frac{M_{1j}}{N_{1j}}, \quad i \ne j, \quad i, j \in {\cal K}\backslash \{1\},
    \label{conds_conli}
\end{equation}
and
\begin{equation}
    \left|\sin(e_i-\tan \eta)\right|\geq \left|\frac{{m_i\lambda}}{2Nd S_i} \right| \quad i \in{\cal K}\backslash \{1\},
    \label{eq:sin_gamma_mul}
\end{equation}
there must exist a pair of solutions for $\beta $ and $\gamma$ to satisfy (\ref{eq:mul_orthog}).
\end{proposition}

Due to the page limit, the proof of Proposition \ref{prop2} is omitted in this paper. It should be noted that in the special case of $K=2$, the condition in (\ref{conds_conli}) can always be satisfied, and the SVO in (\ref{eq:mul_orthog}) is achieved if the angular constraint (\ref{eq:sin_gamma_mul}) holds. For $K>2$, the feasibility region of $\theta_i, i \in \cal K$ is divided into an infinite number of discrete solution sets, each corresponding to a collinear relationship characterized by a rational ratio $\eta$ in (\ref{conds_conli}). Compared with an FOA array, whose feasible region is restricted to a single discrete set, the proposed RAA greatly enlarges the feasible region, thereby increasing the likelihood of satisfying (\ref{eq:mul_orthog}). Nonetheless, as the number of interference directions becomes sufficiently large, it will still be difficult to simultaneously satisfy (\ref{conds_conli}) and (\ref{eq:sin_gamma_mul}).

\subsection{Directional Antenna}
For the directional antenna with the beam pattern given in (\ref{eq:pattern}), we can rewrite the effective SVO in (\ref{orthog_mei}) as 
\begin{equation}\label{newSVO}
\sqrt{g(\mathbf{r}, \epsilon(\theta_k))}\cdot \sqrt{g(\mathbf{r}, \epsilon(\theta_0))}\cdot\mathbf{a}(\mathbf{r},\theta_k)^H \mathbf{a}(\mathbf{r},\theta_0) = 0,    \;    k \in \cal K.
\end{equation}

Furthermore, to achieve the full beam gain at $\theta_0$, the main lobe of each antenna should be aligned to $\theta_0$, leading to
\begin{equation}
    g(\mathbf{r},\epsilon(\theta_0))=g_0\cos^{2p}(\epsilon(\theta_0))=g_0.
    \label{eq:cons_g}
\end{equation}
Hence, (\ref{newSVO}) can be satisfied if either $\mathbf{a}(\mathbf{r},\theta_k)^H \mathbf{a}(\mathbf{r},\theta_0) = 0$ or
\begin{equation}
    g(\mathbf{r}, \epsilon(\theta_k))=0
    \label{gain_0}
\end{equation}
holds for each $i \in \cal K$.

On the other hand, \eqref{eq:cons_g} requires $\cos(\epsilon(\theta_0))=1$. Based on (\ref{eq:sol_ep}) and defining $A = -s_\beta c_{(\gamma+\theta_0)}$ and $B = s_{(\gamma+\theta_0)}$, $\cos(\epsilon(\theta_0))$ can be expressed as $\cos (\epsilon(\theta_0)) = A \sin \alpha + B \cos \alpha$, for which the maximum value and the corresponding $\alpha$ are given by
\begin{equation}
   \cos \epsilon(\theta_0)_{\text{opt}} = \sqrt{A^2 + B^2}, \quad \alpha^* = {\pi}/{2} - \arctan(A/B).
   \label{eq:sol_alpha}
\end{equation}
It follows that unlike isotropic antennas, achieving the effective SVO for directional antennas also depends on $\alpha$. Based on (\ref{eq:sol_alpha}), $\cos(\epsilon(\theta_0))=1$ becomes equivalent to
\begin{equation}
    A^2+B^2=s^2_\beta c^2_{(\gamma+\theta_0)}+s^2_{(\gamma+\theta_0)}=1.
    \label{AB}
\end{equation}
In the following, we still consider two cases with $K=1$ and $K \geq 2$, respectively.
 
\subsubsection{Single Interference Direction ($K=1$)}
In the case of $K=1$, the effective SVO in (\ref{newSVO}) becomes equivalent to 
\begin{equation}
    \sqrt{g(\mathbf{r}, \epsilon(\theta_1))} \cdot\mathbf{a}(\mathbf{r},\theta_1)^H \mathbf{a}(\mathbf{r},\theta_0) = 0,
    \label{C_mei}
\end{equation}
which is satisfied if at least one of (\ref{eq:orthog}) and $g(\mathbf{r}, \epsilon(\theta_1))=0$ is satisfied. We provide the following proposition to characterize the condition for both (\ref{AB}) and (\ref{C_mei}) to hold given a single interference direction $\theta_1$.
\begin{proposition}\label{prop3}
For any RAA with $N>1$, there always exists an ARV solution $\mathbf{r}$ that simultaneously satisfies (\ref{AB}) and (\ref{C_mei}) if and only if
\begin{equation}
    |\theta_0 - \theta_1| \in 
    \left[\arcsin \tfrac{\lambda}{Nd}, 2\pi - \arcsin \tfrac{\lambda}{Nd}\right].
    \label{eq:sep_dir}
\end{equation}
\end{proposition}
\begin{IEEEproof}
As (\ref{C_mei}) is satisfied if (\ref{eq:orthog}) or $g(\mathbf{r}, \epsilon(\theta_1))=0$ is satisfied. We first characterize the condition for both (\ref{eq:orthog}) and (\ref{AB}) to hold. To satisfy (\ref{AB}), after some manipulations, we can simplify it as
\begin{equation}
    {1 - (1-\sin^2\beta)\cos^2(\gamma+\theta_0)}=1,
\end{equation}
which holds if $\sin \beta = \pm 1$ or $\cos(\gamma+\theta_0) = 0$. Taking into account (\ref{eq:cons_beta}) for achieving the SVO in (\ref{eq:orthog}), only the latter can be achieved, leading to
\begin{equation}
    \gamma = {\pi}/{2} - \theta_0 \quad \text{or} \quad {3\pi}/{2} - \theta_0.
    \label{eq:op_gamma}
\end{equation}
By further substituting (\ref{eq:op_gamma}) into (\ref{eq:sol_alpha}), we can derive the corresponding $\alpha$ as $\alpha^* = 0 \ \text{or} \ \pi$.

For the condition in (\ref{eq:orthog}), we go through the same procedures as in the isotropic case and obtain the same condition as in (\ref{eq:cons_beta}). Substituting (\ref{eq:op_gamma}) into (\ref{eq:cons_beta}) yields the solution for $\beta$ as
\begin{equation}
    \cos\beta =  \frac{m_1\lambda}{Nd \sin(\theta_0 - \theta_1)}, 
    \qquad m_1 \in \mathbb{Z}\setminus \mathbb{\tilde{Z}}.
    \label{cos_beta_dir}
\end{equation}
To ensure the right-hand side of (\ref{cos_beta_dir}) lies within $[-1, 1]$, it must hold that
\begin{equation}
    |m_1| \leq ({Nd}/{\lambda}) \cdot \left|\sin ({\theta_0 - \theta_1})\right|.
    \label{eq:cons_m_dir}
\end{equation}
Since $|m_1| \geq 1$ for $N>1$, at least one feasible $m_1$ exists if and only if 
\begin{equation}
    \begin{gathered}
        |\theta_0 - \theta_1| \in 
    \left[\arcsin \tfrac{\lambda}{Nd}, \pi-\arcsin \tfrac{\lambda}{Nd}\right] \\
    \cup \left[ \pi+\arcsin \tfrac{\lambda}{Nd}, 2\pi - \arcsin \tfrac{\lambda}{Nd}\right].
    \label{eq:range_Bd}
\end{gathered}
\end{equation}

Regarding the condition $g(\mathbf{r}, \epsilon(\theta_1))=0$, given that (\ref{AB}) is already met, it can be shown that $g(\mathbf{r}, \epsilon(\theta_1))=0$ is achieved if and only if
\begin{equation}
    |\theta_0-\theta_1|\in [\pi/2, 3\pi/2]
    \label{fallout_cons}
\end{equation}
By taking the union of (\ref{fallout_cons}) and (\ref{eq:range_Bd}), we can obtain the overall angular separation constraint given in (\ref{eq:sep_dir}).
\end{IEEEproof}

Proposition \ref{prop3} reveals that compared with FOA arrays, in addition to the benefits uncovered in the case of isotropic RAAs, the directional RAA offers additional benefits by adjusting its main-lobe direction through array rotation. In contrast, the main lobe of an FOA array is fixed, leading to a significant beam-gain loss at $\theta_0$ under the nulling constraints. Moreover, by comparing (\ref{range_iso_1}) and (\ref{eq:range_Bd}) and leveraging the inequality $\arcsin x \geq 2\arcsin(x/2)$ for $x\geq 0$, it is noted that (\ref{eq:range_Bd}) imposes stricter bounds compared to (\ref{range_iso_1}), which is due to the additional radiation pattern constraint required in (\ref{eq:cons_g}). However, as $x \rightarrow 0$, i.e., the array aperture is sufficiently large, these two bounds become close to each other.

\subsubsection{Multiple Interference Directions ($ K \geq 2$)}
For the case of $K\geq2$, achieving (\ref{newSVO}) requires that either $\mathbf{a}(\mathbf{r},\theta_k)^H \mathbf{a}(\mathbf{r},\theta_0)=0$ or (\ref{gain_0}) be satisfied for each $\theta_k, k \in {\cal K}$, in addition to meeting (\ref{AB}). For each interference direction $\theta_k$, if it satisfies the former condition, i.e., $|\theta_0 - \theta_k| \in \left[\arcsin \tfrac{\lambda}{Nd}, \pi-\arcsin \tfrac{\lambda}{Nd}\right] \cup \left[ \pi+\arcsin \tfrac{\lambda}{Nd}, 2\pi - \arcsin \tfrac{\lambda}{Nd}\right]$ based on \eqref{eq:range_Bd}, the corresponding solution set for $\beta$ can be obtained similarly to (\ref{cos_beta_dir}) by replacing $\theta_1$ therein with $\theta_k$, denoted as
\begin{equation} 
    \begin{gathered}
        {\cal B}_k^{(1)} \triangleq \{\beta | \cos\beta=\frac{m_k\lambda}{Nd \sin(\theta_0 - \theta_k)}, \;  m_k \in \mathbb{Z}\setminus \mathbb{\tilde{Z}}\}.
    \end{gathered}
    \label{beta_set}
\end{equation}
On the other hand, if $\theta_k$ satisfies (\ref{gain_0}), i.e., $|\theta_0-\theta_k|\in [\frac{\pi}{2},\frac{3\pi}{2}]$ based on \eqref{fallout_cons}, no constraints need to be imposed on $\beta$. Hence, the corresponding solution set for $\beta$ is given by ${\cal B}_k^{(2)} \triangleq \{\beta | \beta \in [0, 2\pi)\}$. 

It follows from the above that the overall solution set for $\theta_k$ can be derived as
\begin{equation}
    \mathcal{B}_k =
\begin{cases}
\mathcal{B}_k^{(1)}, \quad |\theta_0-\theta_k|\in {\cal C},\\[2pt]
\mathcal{B}_k^{(2)}, \quad |\theta_0-\theta_k|\in [\frac{\pi}{2},\frac{3\pi}{2}].
\end{cases}
\end{equation}
where $ {\cal C}=\{\theta|\theta \in [\arcsin \tfrac{\lambda}{Nd},\frac{\pi}{2})\cup (\frac{3\pi}{2}, 2\pi-\arcsin \tfrac{\lambda}{Nd})\}$. The optimal $\gamma$ and $\alpha$ to achieve (\ref{AB}) remain the same as in (\ref{eq:op_gamma}). Thus, to achieve (\ref{newSVO}), the intersection of ${\cal B}_k$'s should not be an empty set. We provide the following proposition.
\begin{proposition}\label{prop4}
The full beam gain can be achieved over $\theta_0$ subject to nulling constraints over $\theta_k, k \in {\cal K}$ if and only if $\bigcap_{k=1}^K {{\cal B}_k} \ne \emptyset$.
\end{proposition}

However, the condition provided in Proposition \ref{prop4} is difficult to analyze due to its complex form. To gain insights, we further consider another special case with $K=2$ symmetric interference directions about $\theta_0$, i.e.,
\begin{equation}
    \theta_0 = ({\theta_1 + \theta_2})/{2}, 
    \; \theta_1 = \theta_0 - \Delta_\theta, 
    \; \theta_2 = \theta_0 + \Delta_\theta.
\end{equation}
Then, it can be shown that by setting $m_1=-m_2$, we can always achieve ${\cal B}_1= {\cal B}_2$ and obtain $\beta=\pm\arccos\big((m_1\lambda)/(Nd\sin(\theta_0-\theta_1))\big)$ subject to (\ref{eq:cons_m_dir}). As a result, (\ref{newSVO}) can be fully satisfied.

This section has characterized the conditions to achive (effective) SVO under two specific antenna radiation models. However, if the above conditions are not met or other antenna radiation models are employed, the full beam gain may not be achieved at the direction of $\theta_0$. To characterize the performance of RAA in these cases, we propose a general ARV optimization algorithm in Section \ref{opt}.

\section{Proposed Optimization Algorithm for (P2)}\label{opt}
In this section, we propose a general optimization algorithm to solve (P2), where the 3D angular space is discretized to facilitate a sequential search, jointly with a GS process to avoid low-quality suboptimal solutions.

\subsection{Discrete Sampling and Sequential Update}
We first discretize the 3D angular space $[0,2\pi)^3$ for $(\alpha, \beta,\gamma)$ into $Q^3$ uniformly sampled points, with $Q$ denoting the number of sampling points per dimension. Let $\mathcal{Q} \triangleq \{1,2,\dots,Q\}$ denote the set of grid indices per dimension. Hence, the continuous 3D space can be approximated as a discrete space, and any ARV $\mathbf{r}=[\alpha, \beta, \gamma]^T$ can be approximated as $(\tfrac{2\pi}{Q}\alpha_p,  \tfrac{2\pi}{Q}\beta_p,\tfrac{2\pi}{Q}\gamma_p$), with $\alpha_p, \beta_p, \gamma_p \in \cal Q$. Given the above discretization, (P2) is approximated as
\begin{equation}
    \textrm{(P3)}\quad \max_{\alpha_p,\beta_p,\gamma_p} \; G(\mathbf{r},\mathbf{w}^{\mathrm{ZF}}({\bf{r}}),\theta_0).
\end{equation}
Problem (P3) can be solved via the sequential update algorithm, which sequentially optimizes $\alpha_p,\beta_p$ and $\gamma_p$, while keeping the other two fixed. Denote the ARV in the $(l-1)$-th sequential update round as $\mathbf{r}^{(l-1)} = \big[\alpha_p^{(l-1)},\beta_p^{(l-1)}, \; \gamma_p^{(l-1)}\big]^T$. The update rule of the ARV in the $l$-th round is
\[
\left\{
\begin{aligned}
\alpha_{p}^{(l)} &= 
\arg\max_{\alpha_p \in \mathcal{Q}}
G(\mathbf{r},\mathbf{w}^{\mathrm{ZF}}({\bf{r}}),\theta_0),
\;\beta_p = \beta_p^{(l-1)},\gamma_p = \gamma_p^{(l-1)} \\[-2pt]
\beta_{p}^{(l)} &= 
\arg\max_{\beta_p \in \mathcal{Q}}
G(\mathbf{r},\mathbf{w}^{\mathrm{ZF}}({\bf{r}}),\theta_0),
\;\alpha_p = \alpha_p^{(l)},\gamma_p = \gamma_p^{(l-1)} \\[-2pt]
\gamma_{p}^{(l)} &= 
\arg\max_{\gamma_p \in \mathcal{Q}}
G(\mathbf{r},\mathbf{w}^{\mathrm{ZF}}({\bf{r}}),\theta_0),
\;\alpha_p = \alpha_p^{(l)},\beta_p = \beta_p^{(l)}.
\label{eq:sol_SU}
\end{aligned}
\right.
\]
After that, the ARV is updated as $\mathbf{r}^l = [\alpha_p^l,\beta_p^l,\gamma_p^l]^T,$ which serves as the input for the next round. The process is repeated until a predefined iteration number, denoted as $L$, is reached. However, the above sequential update process may converge to undesired local optimum. To deal with this issue, a GS phase is added to explore the solution space and improve the algorithm robustness, as presented next.

\subsection{GS Phase}

After each round of sequential update, GS is employed to refine the solution in $T$ iterations. Let the solutions output by the GS the $(t-1)$-th iteration be denoted as $\mathcal{E}(t-1) = \{\mathbf{r}_{GS}^{(0)}, \mathbf{r}_{GS}^{(1)}, \dots, \mathbf{r}_{GS}^{(t-1)}\},$ where $\mathbf{r}_{GS}^{(i)} = [\alpha_{p,GS}^i, \;\beta_{p,GS}^i, \; \gamma_{p,GS}^i]^T$ denotes the output of the $i$-th GS iteration, and we set $\mathbf{r}_{GS}^{(0)} = \mathbf{r}^l$.

Next, in the $t$-th iteration, we generate $S^{adj}$ adjacent solutions and $(S-S^{adj})$ random solutions given $\mathbf{r}_{GS}^{(t-1)}$. Each adjacent solution is obtained by shifting either $\alpha_p$ $\beta_p$ or $\gamma_p$ with a certain amount, i.e.,
\[
\left\{
\begin{aligned}
\mathbf{r}_{\alpha,adj}^{(t)} &= 
\big[ \alpha_{p,GS}^{t-1} + j,\ \beta_{p,GS}^{t-1},\ \gamma_{p,GS}^{t-1} \big]^{T}, \\[-2pt]
\mathbf{r}_{\beta,adj}^{(t)} &= 
\big[ \alpha_{p,GS}^{t-1},\beta_{p,GS}^{t-1} + j,\ \gamma_{p,GS}^{t-1} \big]^{T}, \\[-2pt]
\mathbf{r}_{\gamma,adj}^{(t)} &= 
\big[\alpha_{p,GS}^{t-1}, \beta_{p,GS}^{t-1},\ \gamma_{p,GS}^{t-1} + j \big]^{T},
\label{eq:set_B}
\end{aligned}
\tag{54}
\right.
\]
with $j \in [-J,J]$, where $J$ denotes the maximum shifting amount. This gives rise to $S^{adj}=6J$ adjacent solutions in total. Let $\mathcal{B}(t)$ denote the set of all adjacent solutions in the $t$-th GS iteration. The remaining $(S-S^{adj})$ solutions are randomly generated within the 3D angular domain, and we define $\mathcal{D}(t)$ as the set of all random solutions in the $t$-th GS iteration.

After that, one candidate solution is selected from the set $\mathcal{B}(t)\cup\mathcal{D}(t)$ according to the following probability distribution:
\setcounter{equation}{54}
\begin{equation}
    P_s^{(t)} = \frac{\exp\big(\mu G({\mathbf{r},\mathbf{w}^{\mathrm{ZF}}({\bf{r}})},\theta_0)\big)}
    {\sum_{\mathbf{r}'\in \mathcal{B}(t)\cup \mathcal{D}(t)} \exp\big(\mu G({\mathbf{r}',\mathbf{w}^{\mathrm{ZF}}({\bf{r}}')},\theta_0)\big)},
    \label{eq:probablity}
\end{equation}
where $\mu>0$ is a predefined parameter. The selected solution $\mathbf{r}_{GS}^{(t)}$ is then added to the history set as $\mathcal{E}(t) = \mathbf{r}_{GS}^{(t)} \cup \mathcal{E}(t-1)$ to finalize the $t$-th iteration. After $T$ iterations, we select the best solution among $\mathcal{E}(T)$ as
\begin{equation}
    \mathbf{r}^l = \arg\max_{\mathbf{r}\in\mathcal{E}(T)} G(\mathbf{r},\mathbf{w}^{\mathrm{ZF}}({\bf{r}}),\theta_0),
    \label{eq:update}
\end{equation}
and input (\ref{eq:update}) to the $(l+1)$-th round of sequential update.

Building upon the above procedure, it should be noted that the sequential update process always guarantees a nondecreasing objective value of (P2) over multiple rounds. Moreover, as $\mathbf{r}^l$ is included in the solution set $\mathcal{E}(T)$ in the GS, the GS always output a solution with its performance no worse than $r^l$. Hence, the algorithm is ensured to converge. Regarding the complexity order, note that each sequential update round and GS iteration incur a complexity order of $\mathcal{O}(Q)$, and ${\cal {O}}(S)$, respectively. Thus, the overall complexity of the algorithm is $\mathcal{O}(LQ+ST),$ which is linear and practically efficient.

\begin{figure}[!t]
  \centering
  \begin{subfigure}[t]{0.5\linewidth}
    \centering
    \includegraphics[width=\linewidth]{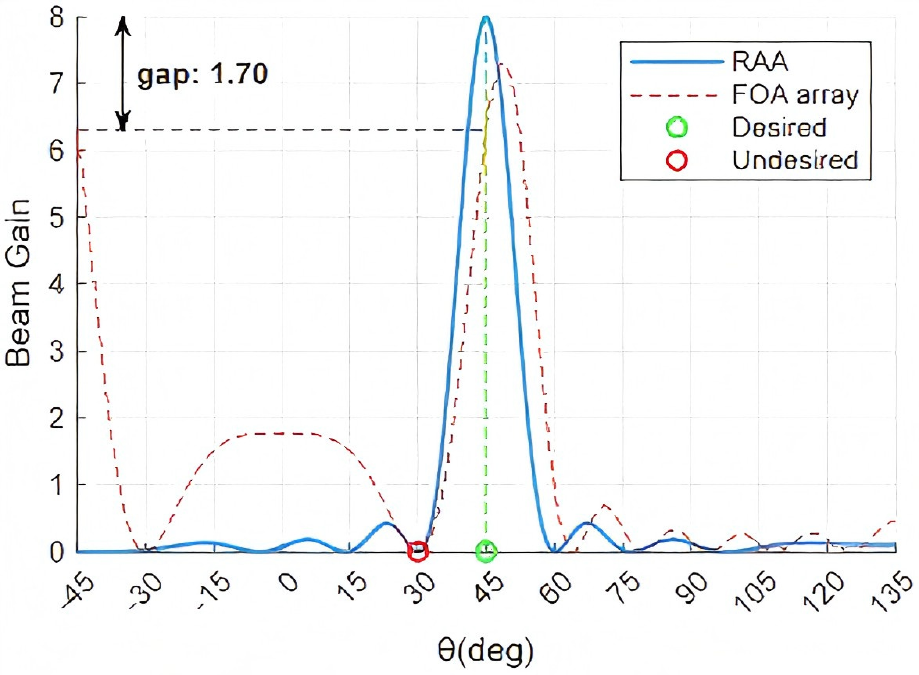}
    \caption{Isotropic $(K=1)$}
  \end{subfigure}\hfill%
  \begin{subfigure}[t]{0.5\linewidth}
    \centering
    \includegraphics[width=\linewidth]{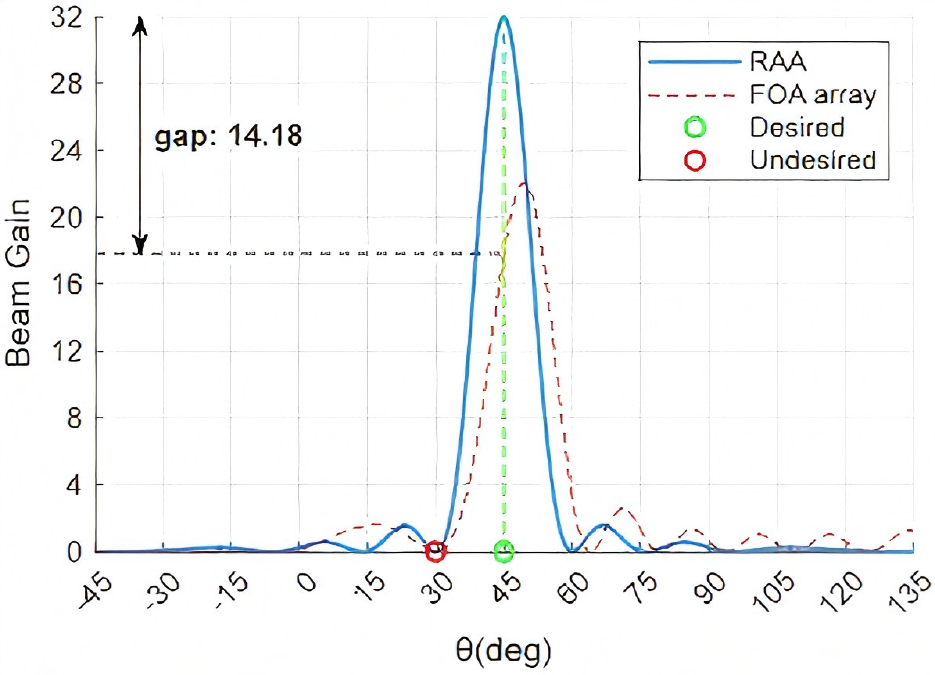}
    \caption{Directional $(K=1)$}
  \end{subfigure}

  \vspace{0.8em}

  \begin{subfigure}[t]{0.5\linewidth}
    \centering
    \includegraphics[width=\linewidth]{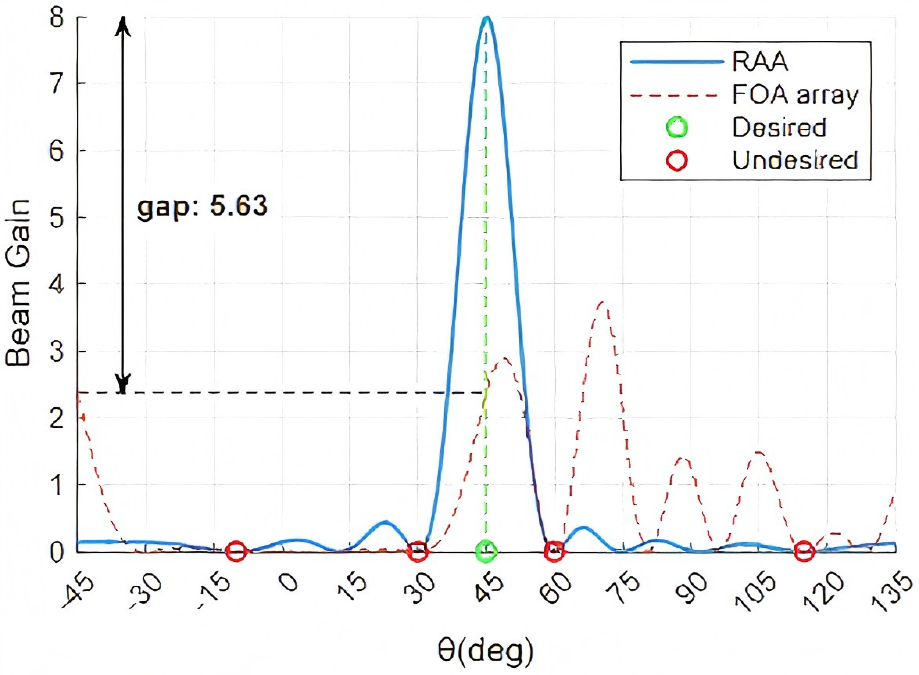}
    \caption{Isotropic $(K=4)$}
  \end{subfigure}\hfill%
  \begin{subfigure}[t]{0.5\linewidth}
    \centering
    \includegraphics[width=\linewidth]{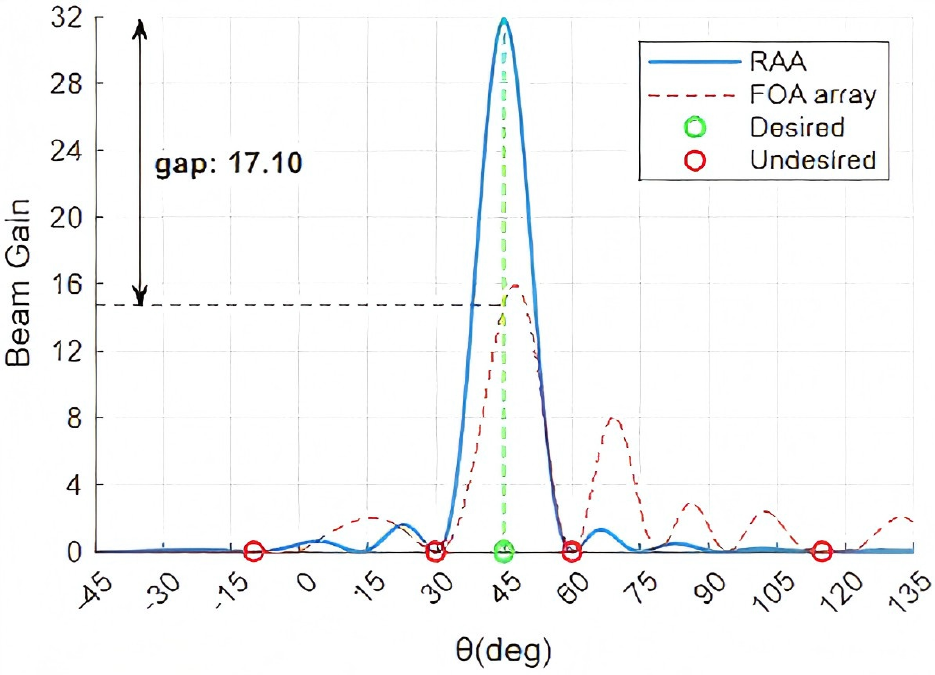}
    \caption{Directional $(K=4)$}
  \end{subfigure}
  \caption{Comparisons of beam pattern between RAA and FOA arrays.}
  \vspace{-12pt}
  \label{fig:beam-compare}
\end{figure}

\vspace{-3pt}
\section{Numerical Results}
In this section, we present numerical results to evaluate the null-steering performance of the proposed algorithm. The number of antennas is fixed at $N=8$ with half-wavelength spacing. We consider the same antenna radiation pattern models as in Section III. For directional antennas, we adopt the cosine gain model with $p=1/2$ in (\ref{eq:pattern}). The desired direction is fixed as $\theta_0 = 45^\circ$. In the proposed sequential update algorithm, the number of sampling points per dimension is set to $Q = 360$, while each GS iteration generates $S = 36$ candidates. The maximum iterations of sequential update and GS are set to $L = 5$ and $T = 50$, respectively.

Fig.\,\ref{fig:beam-compare} compares the beam patterns of the proposed RAA and the conventional FOA arrays in the cases of $K=1$ and $K=4$. For $K=1$, we set $\theta_1=30^\circ$; while for $K=4$, we set $\theta_1 = -10^\circ, \, \theta_2 = 30^\circ, \, \theta_3 = 60^\circ$ and $\theta_4 = 115^\circ$. In the directional case, the maximum array gain equals the isotropic full-array gain scaled by the directional factor $g_e(\epsilon_0)=2(2p+1)=4$, i.e., $4N$. It is observed that for FOA arrays, as the ZF beamforming enforces strict nulls, there exists a main-lobe misalignment at $\theta_0$ for both isotropic and directional antenna patterns. Particularly, the beam-gain loss becomes significantly high (over 3 dB) for $K=4$. In contrast, the proposed RAA exploits array rotation to align the desired steering vector with the main lobe while keeping orthogonality to all interference directions. For $K=1$, since $|\theta_0-\theta_1|$ falls within the interval provided in (\ref{range_iso_1}) and (\ref{eq:sep_dir}), the RAA achieves perfect effective SVO. For $K=4$, it attains near-full beam gains of $7.99$ and $31.80$ in the isotropic and directional cases, respectively, confirming the effectiveness of the proposed optimization algorithm. As an additional merit, it is also observed from Fig. \,\ref{fig:beam-compare} that the RAA can dramatically reduce the side-lobe levels compared to FOA arrays. In Fig.\,\ref{fig:beam-compare} (c), the side-lobe level of the FOA array becomes even higher than its main-lobe level.

\begin{figure}[!t]
  \centering
  \begin{subfigure}[t]{0.5\linewidth}
    \centering
    \includegraphics[width=\linewidth,]{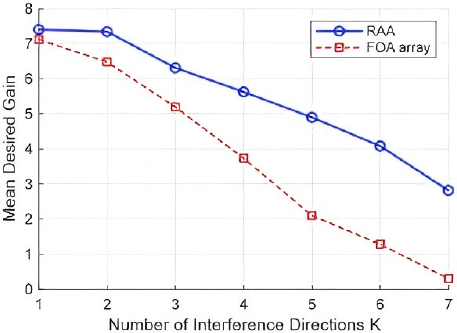}
    \caption{Isotropic}
  \end{subfigure}\hfill%
  \begin{subfigure}[t]{0.5\linewidth}
    \centering
    \includegraphics[width=\linewidth]{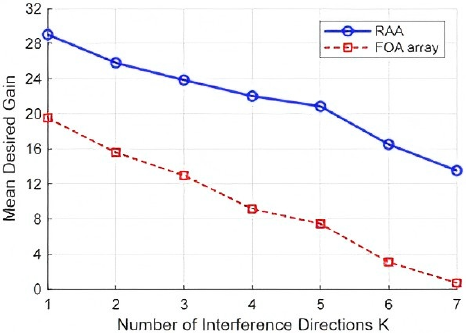}
    \caption{Directional}
  \end{subfigure}
  \caption{Beam gain over the desired direction versus number of interference directions.}
  \vspace{-9pt}
  \label{fig:MonteCarlo}
\end{figure}

In Fig.\,\ref{fig:MonteCarlo}, we plot the beam gain over the desired direction versus the number of interference directions $K$ under both isotropic and directional antenna patterns. The interference directions are randomly generated between $0$ and $2\pi$, and the results are averaged over 100 Monte Carlo trials. As observed from Fig.\,\ref{fig:MonteCarlo}, the desired gain decreases as $K$ increases in both isotropic and directional cases due to the stricter nulling constraints. Nevertheless, the proposed RAA consistently achieves a higher beam gain than the conventional FOA arrays across all values of $K$. In particular, while the beam gain of the FOA array rapidly drops to nearly zero as $K$ approaches $7$, the RAA exhibits a much slower degradation of beam gain, thereby achieving superior null-steering performance.

\section{Conclusion}
In this paper, we investigated the null-steering optimization for an RAA, aiming to jointly optimize the 3D ARV of the RAA to maximize the beam gain over the desired direction under ZF beamforming. To gain insights, we derived the (effective) SVO conditions in several special cases including both isotropic and directional antenna radiation patterns. For other general cases, we developed a Gibbs sampling-enhanced sequential update algorithm to obtain a high-quality suboptimal ARV solution. Simulations confirmed that the RAA significantly outperforms the conventional FOA array.

\end{document}